\documentclass[11pt]{article}

\usepackage{amsmath,amssymb,bm}
\usepackage{graphicx,color,overpic}
\graphicspath{{figures/}}

\usepackage[a4paper,left=2.1cm,right=2.1cm,top=2.2cm,bottom=2.2cm]{geometry}
\newcommand\mymatrix[1]{\bm{\mathrm{#1}}}
\usepackage{natbib,url}
\usepackage{lineno}
\usepackage{multirow,bigstrut}
\definecolor{dgreen}{rgb}{0,.6,0}

\bibpunct{[}{]}{;}{a}{,}{,}

\newlength\figwidth
\newlength\imagewidth
\newtheorem{fact}{Fact}

\begin{document}

\title{\bf Breaking a Chaotic Cryptographic Scheme Based on Composition Maps%
\author{Chengqing Li\textsuperscript{1}, David Arroyo\textsuperscript{2}, and Kwok-Tung
Lo\textsuperscript{1}\\
\textsuperscript{1} Department of Electronic and Information
Engineering, \\The Hong Kong
Polytechnic University, Hong Kong, China\\[0.5em]
\textsuperscript{2}Instituto de F\'{\i}sica Aplicada, Consejo
Superior de Investigaciones Cient\'{\i}ficas, \\Serrano 144, 28006
Madrid, Spain}}


\maketitle

\begin{abstract}
Recently, a chaotic cryptographic scheme based on composition maps
was proposed. This paper studies the security of the scheme and
reports the following findings: 1) the scheme can be broken by a
differential attack with $6+\lceil\log_L(MN)\rceil$
chosen-plaintext, where $MN$ is the size of plaintext and $L$ is the
number of different elements in plain-text; 2) the scheme is not
sensitive to the changes of plaintext; 3) the two composition maps
do not work well as a secure and efficient random number source.
\end{abstract}

\section{Introduction}

The development of information technology makes the transmission of
digital data is carried out more and more frequently over all kinds
of channels. Meanwhile, the security of digital data become more and
more important. So, the demand of secure and fast encryption schemes
become urgent. Due to the subtle similarities between cryptography
and chaos, a great number of chaotic encryption schemes have been
proposed in the past decade
\cite{YaobinMao:CSF2004,Pareek:ImageEncrypt:IVC2006,Flores:EncryptLatticeChaos06,
Tong:ImageCipher:IVC07,Wong:ChaosEncrypt:IEEETCASII,Pareek:CNSNS2009}.
However, most of them have been found to be insecure in different
extents from the view point of modern cryptography
\cite{AlvarezLi:Cryptanalysis:PLA2005,LiShujun:ReturnMapAttack:IJBC2006,
Li:AttackDSEA2006,Li:AttackHDSP2006,Li:AttackingBitshiftXOR2007,Li:AttackingRCES2004,
ArroyoLi:Chaos2008,LiLi:IVC2009b,Li:ProbabilisticEncryption:IJBC03,Li:BreakNoiseCommunication:Chaos05,Li:BreakYTSCipher:TCASII04}.
As for how to evaluate the security of a chaotic cryptographic
scheme, please refer to \cite{LiShujun:Rules:IJBC2006}.

In general, the usage of chaos in designing encryption scheme can be
classified as three categories: 1) generating pseudo-random number
sequence, which is then used to determine position permutation; 2)
generating pseudo-random bit sequence, which is then used to
determine combination and/or composition of some basic encryption
operations; 3) generating ciphertext directly when the data of
plaintext is assigned as the initial condition or control parameter
of a chaotic map. In
\cite{Akhavan:LNCS06,Akhshani:ICIP06,Behnia:IJBC08}, the possible
application of composition of polynomial chaotic maps in designing
encryption scheme was discussed. In this case, two composite
polynomial chaotic maps are used to determine the position
permutation and composition of basic encryption operations
respectively. Since the schemes proposed in
\cite{Akhavan:LNCS06,Akhshani:ICIP06} are preliminary version of the
one proposed in \cite{Behnia:IJBC08}, this paper only focuses on the
security of the latter. With our study, the following security problems are found:
1) the scheme can be broken with a differential attack; 2) the
scheme is not sensitive with respect to the changes of plaintext; 3)
the randomness of the pseudo-random number sequences generated by
the two composition maps is weak.

The rest of this paper is organized as follows.
Section~\ref{sec:scheme} describes the chaotic cryptographic scheme
briefly. A comprehensive cryptanalysis on the scheme is presented in
Sec.~\ref{sec:Cryptanalysis}. The last section concludes this paper.

\section{The Encryption Scheme Under Study}
\label{sec:scheme}

In \cite{Behnia:IJBC08}, the structure of plaintext is not specified
precisely. Without loss of generality, the plaintext here is denoted
by a 2-D byte array of size $M\times N$ (height$\times$width),
$\mymatrix{I}=\{I(i,j)\}_{1\leq i\leq M \atop 1\leq j\leq N}$ and
the corresponding ciphertext by $\mymatrix{I}'=\{I'(i,j)\}_{1\leq
i\leq M \atop 1\leq j\leq N}$. The plaintext is considered as a 1D
signal $\{I(k)\}_{k=1}^{MN}$ by scanning it in a raster order. Then,
the chaotic cryptographic scheme can be described as follows
\footnote{To make the presentation more concise and complete, some
notations in the original paper are modified, and some details about
the scheme are supplied and/or corrected also.}.

\begin{itemize}
  \item \textit{Secret key:} three sets of initial condition and
  control parameter of Eq.~(\ref{eq:fx}), $(x_0, \alpha_1,
  \alpha_2)$, $(x_0', \alpha_1', \alpha_2')$, $(x_0^*, \alpha_1^*,
  \alpha_2^*)$,  one set of initial condition and
  control parameter of Eq.~(\ref{eq:gx}), $(y_0, \alpha_3,
  \alpha_4)$, and a secret number $S\in \{0, \cdots, 255\}$.

  \begin{equation}
 f(x)=\frac{1}{{\alpha_{{2}}}^{2}}\tan^{2} \left( 5\,\arctan
\left( {\frac {  \tan ( 3\,\arctan ( \sqrt{x} )) }{{ \alpha_{{1}}}}}
\right)  \right). \label{eq:fx}
\end{equation}

\begin{equation}
g(y)= \frac{1}{{\alpha_{4}}^{2}} \cot^{2} \left( 8\,\arctan
\left(\alpha_3 \tan \left( 4\,\arctan \left( {\frac {1}{\sqrt {y}}}
\right)\right)
  \right)\right).
 \label{eq:gx}
\end{equation}

  \item \textit{Initialization:}
\begin{itemize}
\item Iterate the map Eq.~(\ref{eq:fx}) $MN$ times to obtain three
states sequences, $\{\psi_1(k)\}_{k=1}^{MN}$,
$\{\psi_2(k)\}_{k=1}^{MN}$,
 $\{\psi_4(k)\}_{k=1}^{MN}$, under the three sets of
 initial condition and control parameter, $(x_0, \alpha_1,
  \alpha_2)$, $(x_0', \alpha_1', \alpha_2')$, $(x_0^*, \alpha_1^*,
  \alpha_2^*)$, respectively;

\item Iterate the map Eq.~(\ref{eq:gx}) $MN$ times to obtain a
states sequence $\{\psi_3(k)\}_{k=1}^{MN}$ under initial condition
and control parameter $(y_0, \alpha_3, \alpha_4)$;

\item Generate four pseudo-random number sequences, $\{\phi_1(k)\}_{k=1}^{MN}$,
$\{\phi_2(k)\}_{k=1}^{MN}$, $\{\phi_3(k)\}_{k=1}^{MN}$,
$\{\phi_4(k)\}_{k=1}^{MN}$, as follows:
$\phi_1(k)=\lfloor\psi_1(k)\cdot 10^{14}\rfloor \bmod M$,
$\phi_2(k)=\lfloor\psi_2(k)\cdot 10^{14}\rfloor \bmod N$,
$\phi_3(k)=\lfloor\psi_3(k)\cdot 10^{14}\rfloor \bmod 256$, and
$\phi_4(k)=\lfloor\psi_4(k)\cdot 10^{14}\rfloor \bmod 256$.
\end{itemize}

\item \textit{Encryption:}
  \begin{itemize}
    \item \textit{Permutation:} for $k=1 \sim MN$,
    swap the positions of two bytes $I(k)$ and $I(\phi_1(k)\cdot N+\phi_2(k))$.
    Denote the permuted plaintext with $\mymatrix{I}^*=\{I^*(i,j)\}_{1\leq i\leq M \atop 1\leq j\leq N}$.

    \item \textit{Confusion I:}
    for $k=1 \sim MN$,
    \begin{equation}
    I^\star(k)=\phi_3(k)\oplus (I^*(k)\dotplus \phi_3(k))\oplus
   I^\star(k-1),\label{eq:MaskEncryptI}
    \end{equation}
    where $I^\star(0)=S$, and $x\dotplus y=(x+y)\bmod 256$.

    \item \textit{Confusion II:}
    for $k=1 \sim MN$,
    \begin{equation}
    I'(k)= I^\star(k)\oplus \phi_4(k).
     \end{equation}

\end{itemize}

\item \textit{Decryption:}
The decryption approach is similar to the encryption one except that
the main three encryption steps and the swap operations in the
permutation step are carried out in a reverse order, and
Eq.~(\ref{eq:MaskEncryptI}) is replaced by the following function.

\begin{equation}
    I^*(k)=((I^\star(k)\oplus I^\star(k-1)\oplus\phi_3(k))-\phi_3(k)+256)\bmod
    256.
    \end{equation}

\end{itemize}

\section{Cryptanalysis}
\label{sec:Cryptanalysis}

\subsection{Differential Attack}

Differential attack is an attack to recover the information about
secret key and/or plaintext by analyzing the evolution of
differences when some pairs of plaintexts are encrypted with the
same secret key. In \cite[Sec.~5.4]{Behnia:IJBC08}, the authors
claimed the encryption scheme under study can withstand differential
attack effectively. However, we find that the scheme can be broken by this
attack easily with the following steps.

\begin{itemize}
\item \textit{Breaking Confusion I:}

If two plaintexts, $\mymatrix{I}_1=\{I_1(k)\}_{k=1}^{MN}$ and
$\mymatrix{I}_2=\{I_2(k)\}_{k=1}^{MN}$, are encrypted by the same
secret key, one has the following equality.
\begin{eqnarray}
    I'_1(k)\oplus I'_2(k) & = & I_1^\star(k)\oplus \phi_4(k)\oplus I_2^\star(k)\oplus
    \phi_4(k) \nonumber\\
    & = & I^\star_1(k)\oplus I^\star_2(k) \nonumber\\
    & = & \phi_3(k)\oplus (I_1^*(k)\dotplus \phi_3(k))\oplus
   I^\star_1(k-1)\oplus   \\
&  &   \phi_3(k) \oplus (I^*_2(k)\dotplus \phi_3(k))\oplus
   I^\star_2(k-1) \nonumber\\
   & = & (I_1^*(k)\dotplus \phi_3(k))\oplus (I_2^*(k)\dotplus
   \phi_3(k))\oplus(I^\star_1(k-1) \oplus I^\star_2(k-1))\nonumber\\
   & = & (I_1^*(k)\dotplus \phi_3(k))\oplus (I_2^*(k)\dotplus
   \phi_3(k))\oplus(I'_1(k-1) \oplus I'_2(k-1))
\end{eqnarray}

Furthermore, if the plaintexts, $\mymatrix{I}_1$ and
$\mymatrix{I}_2$, are chosen of fixed value, one has
\begin{equation}
 (I'_1(k)\oplus I'_2(k)) \oplus(I'_1(k-1) \oplus I'_2(k-1))=(I_1(k)\dotplus \phi_3(k))\oplus (I_2(k)\dotplus
   \phi_3(k))\label{eq:solvephi3}
\end{equation}

Since the left part of the above equation, $I_1(k)$ and $I_2(k)$ are
known, Eq.~(\ref{eq:solvephi3}) can be simplified as the following
equation.
\begin{equation}
y=(a\dotplus x) \oplus (b\dotplus x), \label{eq:equivalentfunction}
\end{equation}
where $a, b, x \in\{0, 1, \cdots 255\}.$

It has been verified by computer that a set $\{x, x\oplus 128\}$ can
be determined uniquely with three different sets of $(a, b)$, e.g.
(9, 127), (1, 52), (33, 65). From Fact~\ref{corollary:xor128}, one
can see that $\phi_3(k)$ and $\phi_3(k)\oplus 128$ are equivalent
with respect to the encryption.

\begin{fact}\label{corollary:xor128}
$\forall\ a,b\in \mathbb{Z}$, $(a\oplus 128)\dotplus b=(a\dotplus
b)\oplus 128$.
\end{fact}

\item \textit{Breaking Confusion II:}

After $\{\phi_3(k)\}_{k=1}^{MN}$ has been broken, only the step
\textit{Confusion II} is left for a plaintext of fixed value,
$\mymatrix{I}_1$, $I_1^\star(k)$ can be determined. Then, one has
\begin{equation}
    \phi_4(k)= I_1'(k)\oplus I_1^\star(k),
\end{equation}
for $k=1\sim MN$.

\item \textit{Breaking Permutation:}

After the steps \textit{Confusion I} and \textit{Confusion II} have
been broken, only the step \textit{permutation} is left. As shown in
\cite{Li:AttackingPOMC2004}, any permutation-only cryptographic
scheme can be broken with only $O\left(\log_L(MN)\right)$
known/chosen plain-texts, where $L$ is the number of different
element in plain-text.
\end{itemize}

\begin{figure}[!htb]
\centering
\begin{minipage}[t]{0.75\imagewidth}
\raggedright
\includegraphics[width=0.75\imagewidth]{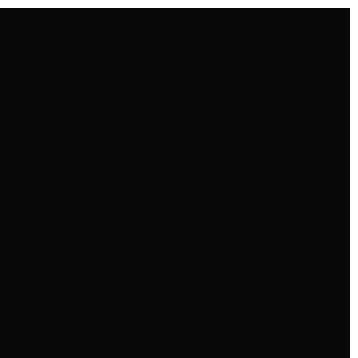}
a)
\end{minipage}
\begin{minipage}[t]{0.75\imagewidth}
\raggedright
\includegraphics[width=0.75\imagewidth]{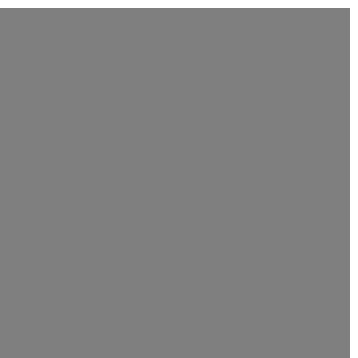}
b)
\end{minipage}
\begin{minipage}[t]{0.75\imagewidth}
\raggedright
\includegraphics[width=0.75\imagewidth]{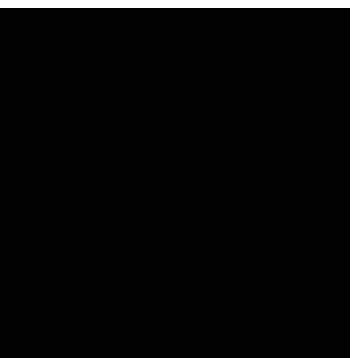}
c)
\end{minipage}
\begin{minipage}[t]{0.75\imagewidth}
\raggedright
\includegraphics[width=0.75\imagewidth]{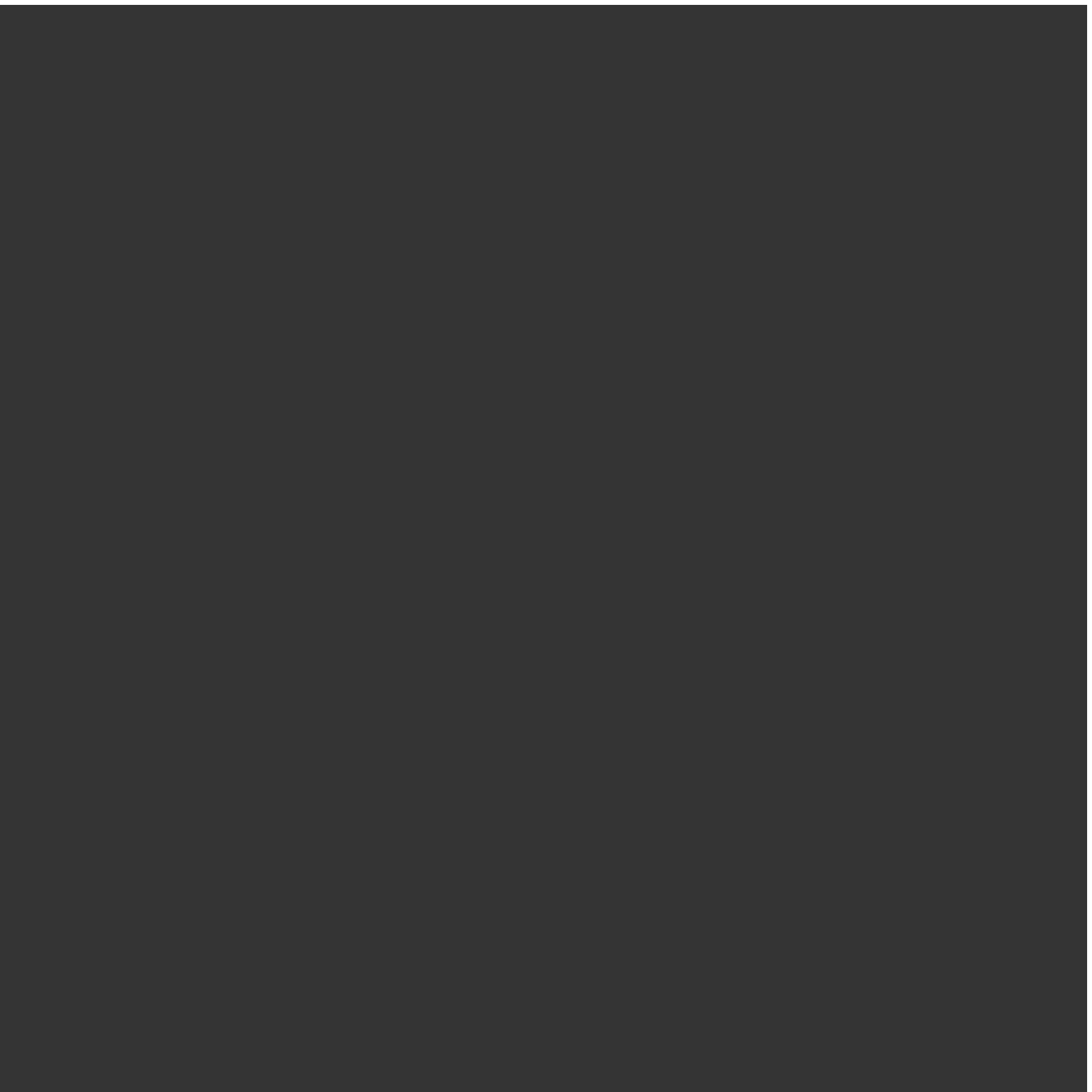}
d)
\end{minipage}
\begin{minipage}[t]{0.75\imagewidth}
\raggedright
\includegraphics[width=0.75\imagewidth]{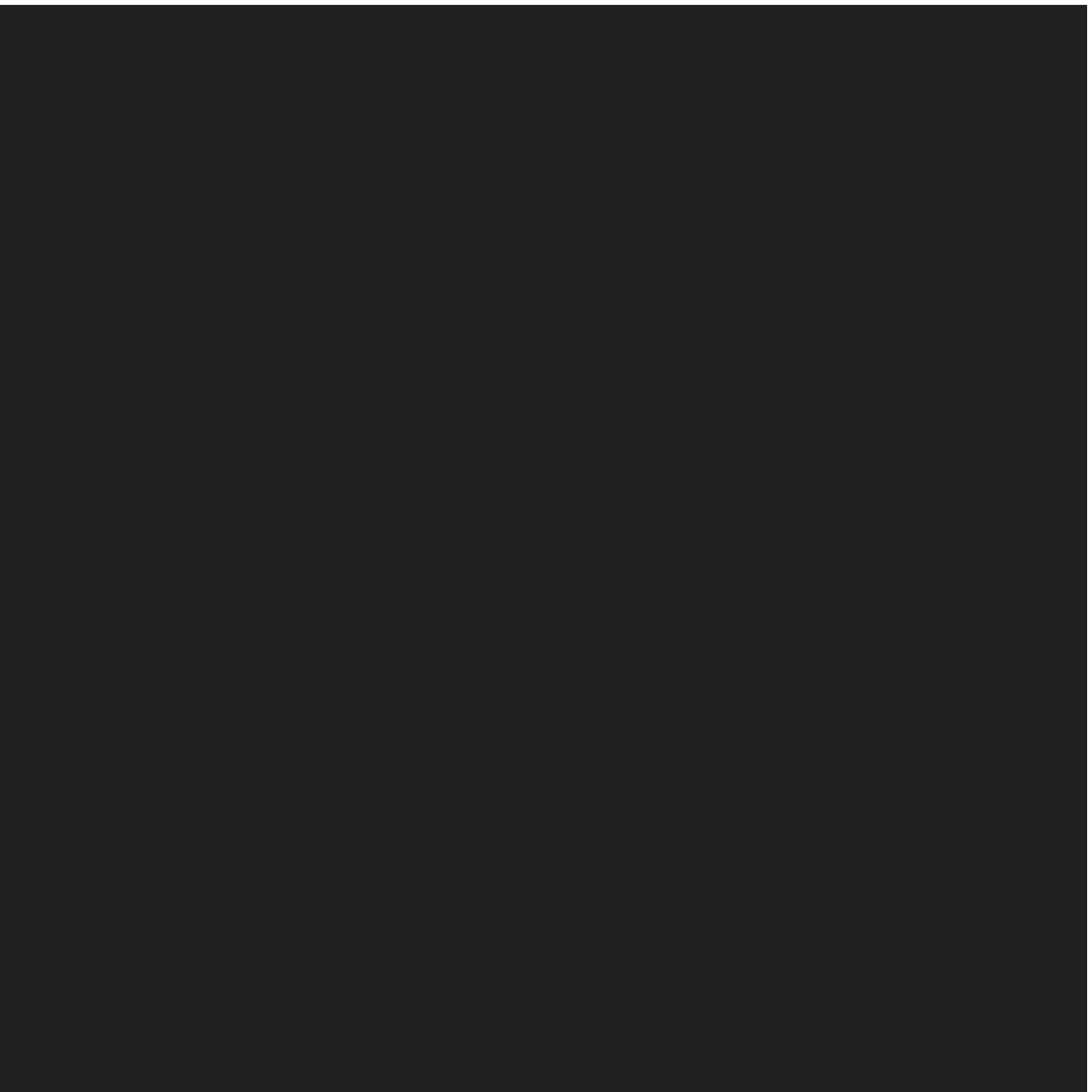}
e)
\end{minipage}
\begin{minipage}[t]{0.75\imagewidth}
\raggedright
\includegraphics[width=0.75\imagewidth]{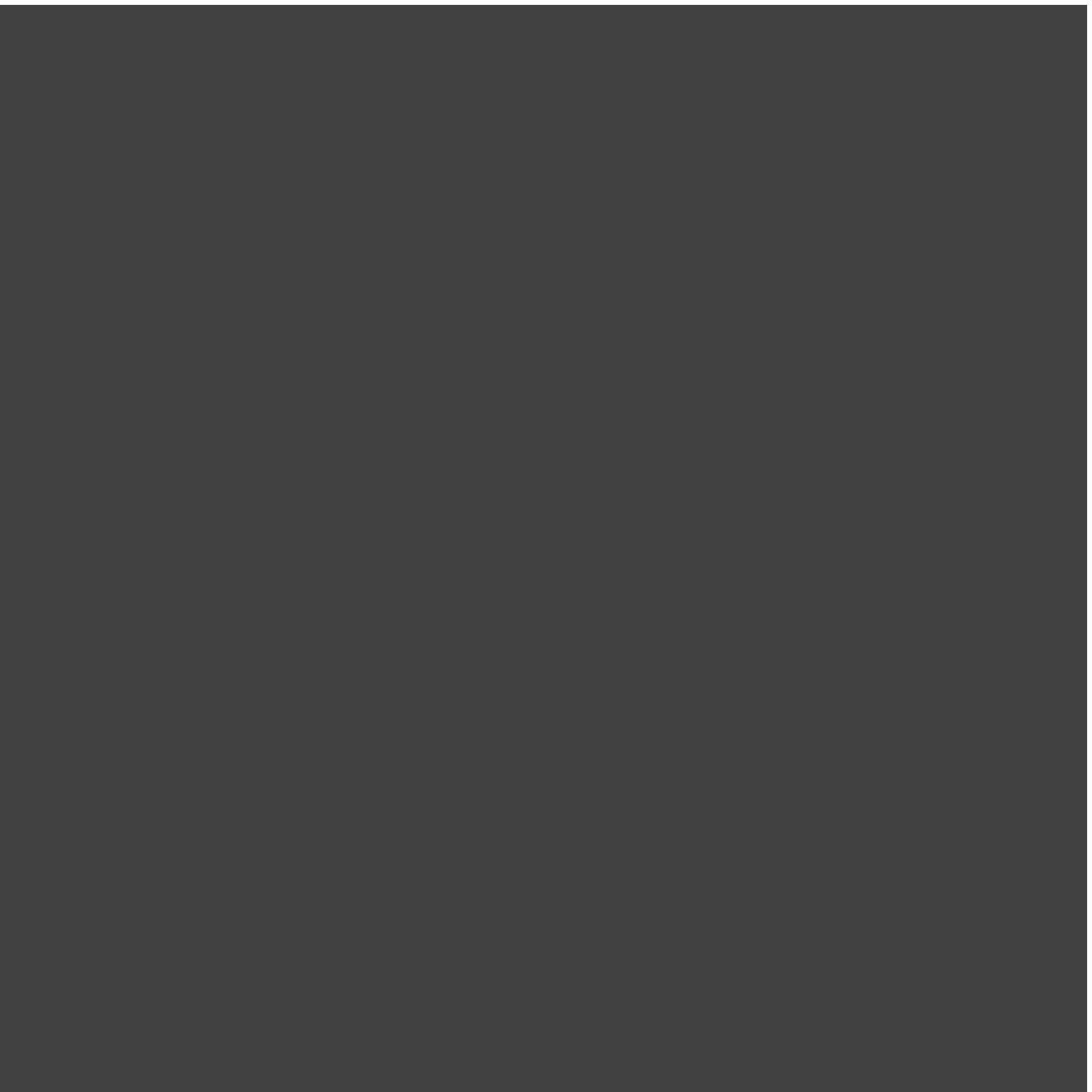}
f)
\end{minipage}
\caption{Six chosen plain-images for breaking Confusion I.}
\label{figure:BreakingConfusion}
\end{figure}

\begin{figure}[!htb]
\centering
\begin{minipage}[t]{0.75\imagewidth}
\raggedright
\includegraphics[width=0.75\imagewidth]{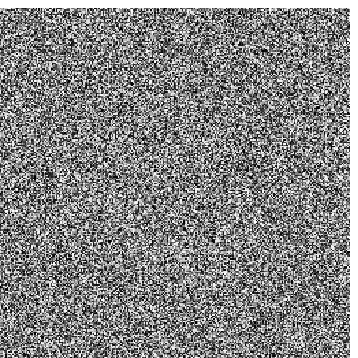}
a)
\end{minipage}
\begin{minipage}[t]{0.75\imagewidth}
\raggedright
\includegraphics[width=0.75\imagewidth]{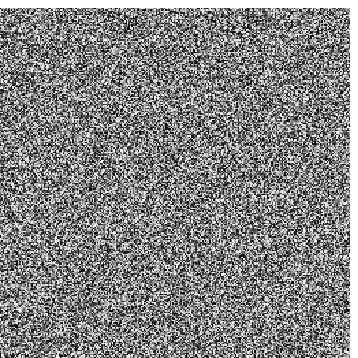}
b)
\end{minipage}
\begin{minipage}[t]{0.75\imagewidth}
\raggedright
\includegraphics[width=0.75\imagewidth]{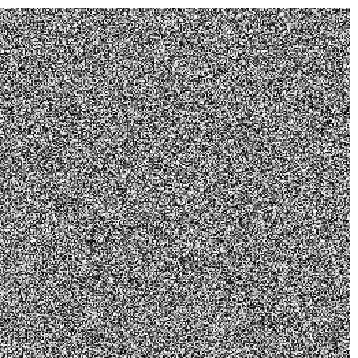}
c)
\end{minipage}
\begin{minipage}[t]{0.75\imagewidth}
\raggedright
\includegraphics[width=0.75\imagewidth]{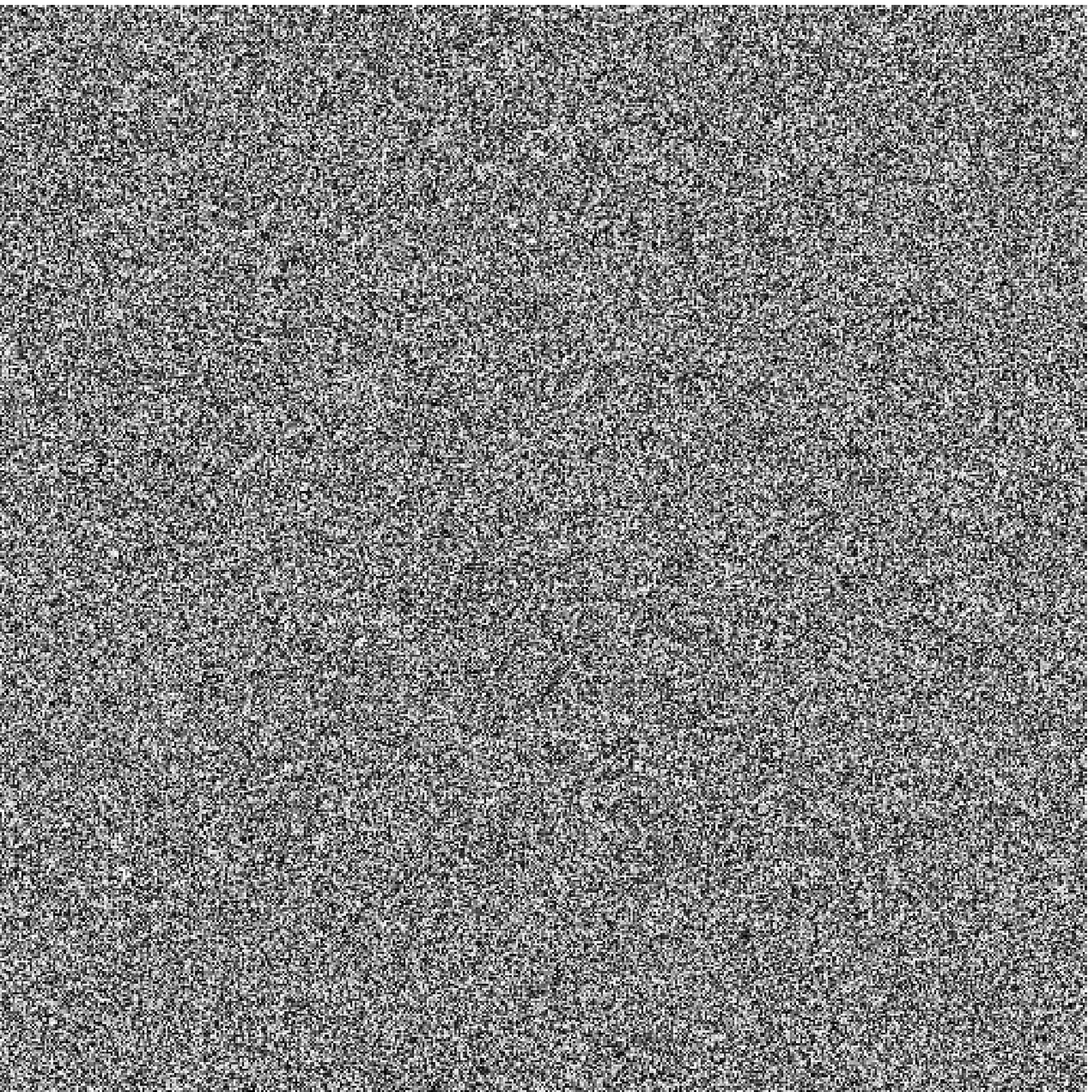}
d)
\end{minipage}
\begin{minipage}[t]{0.75\imagewidth}
\raggedright
\includegraphics[width=0.75\imagewidth]{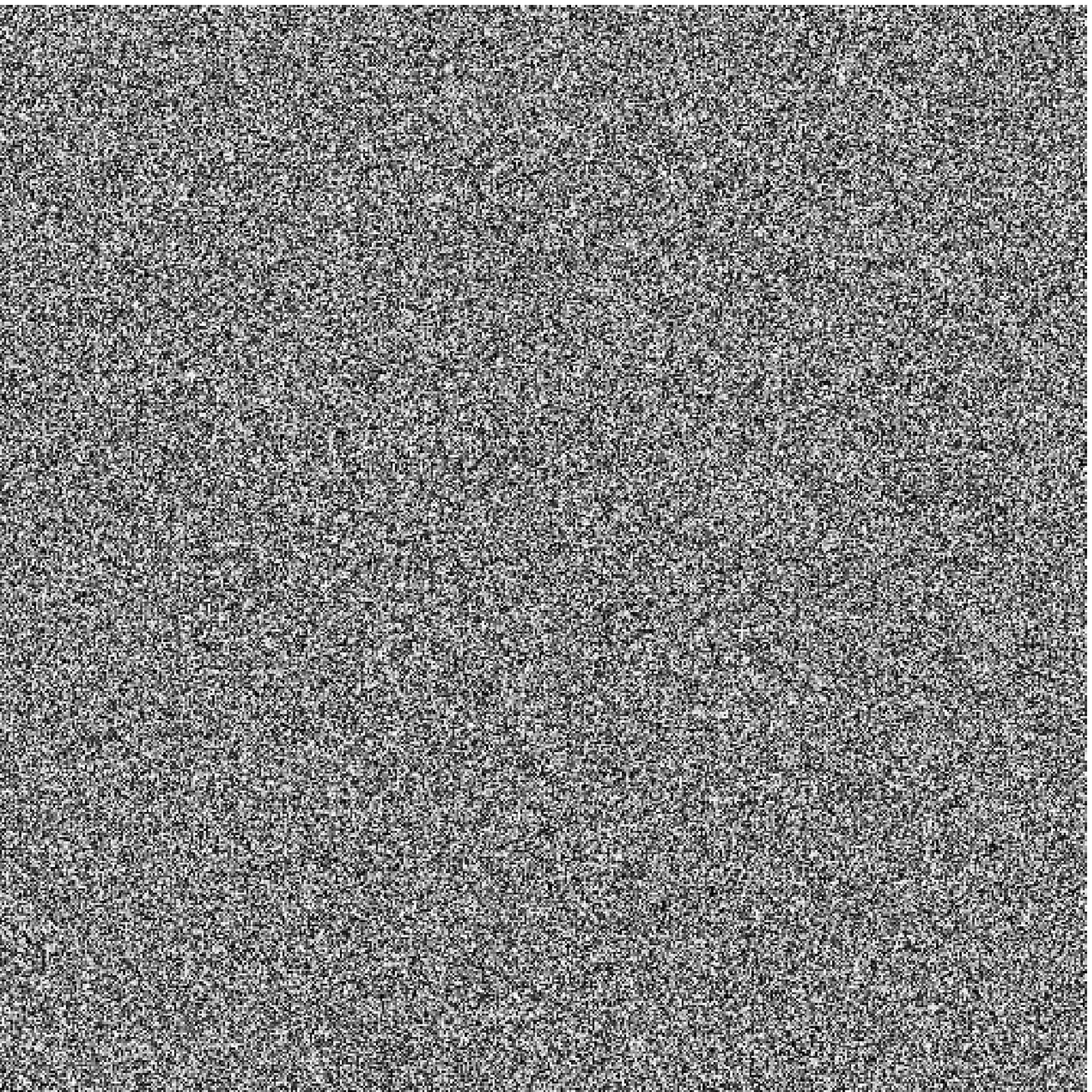}
e)
\end{minipage}
\begin{minipage}[t]{0.75\imagewidth}
\raggedright
\includegraphics[width=0.75\imagewidth]{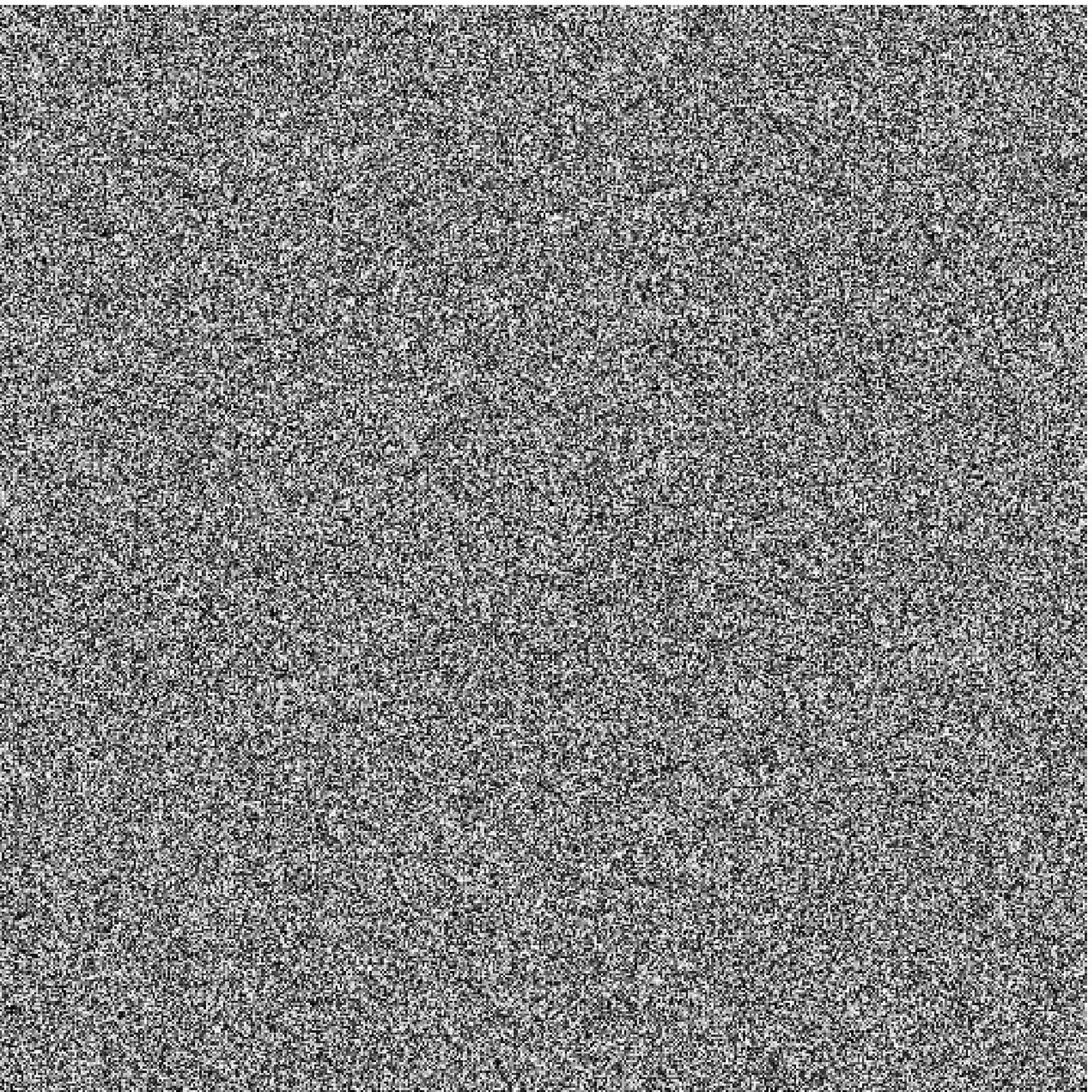}
f)
\end{minipage}
\caption{The cipher-images of the above six chosen plain-images
shown in Fig.~\ref{figure:BreakingConfusion}.}
\label{figure:BreakingConfusion2}
\end{figure}

\begin{figure}[!htb]
\centering
\begin{minipage}[t]{0.75\imagewidth}
\raggedright
\includegraphics[width=0.75\imagewidth]{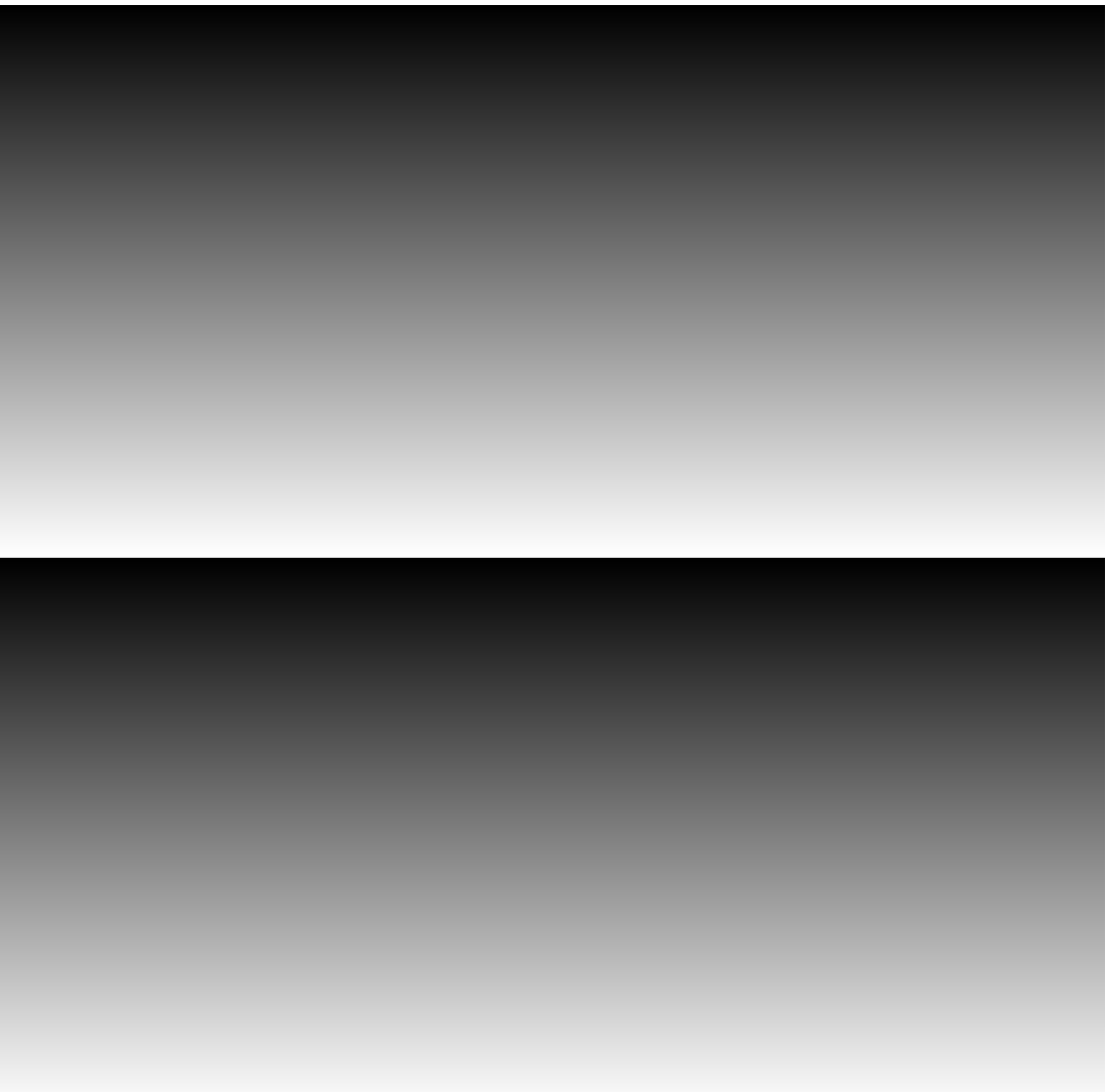}
a)
\end{minipage}
\begin{minipage}[t]{0.75\imagewidth}
\raggedright
\includegraphics[width=0.75\imagewidth]{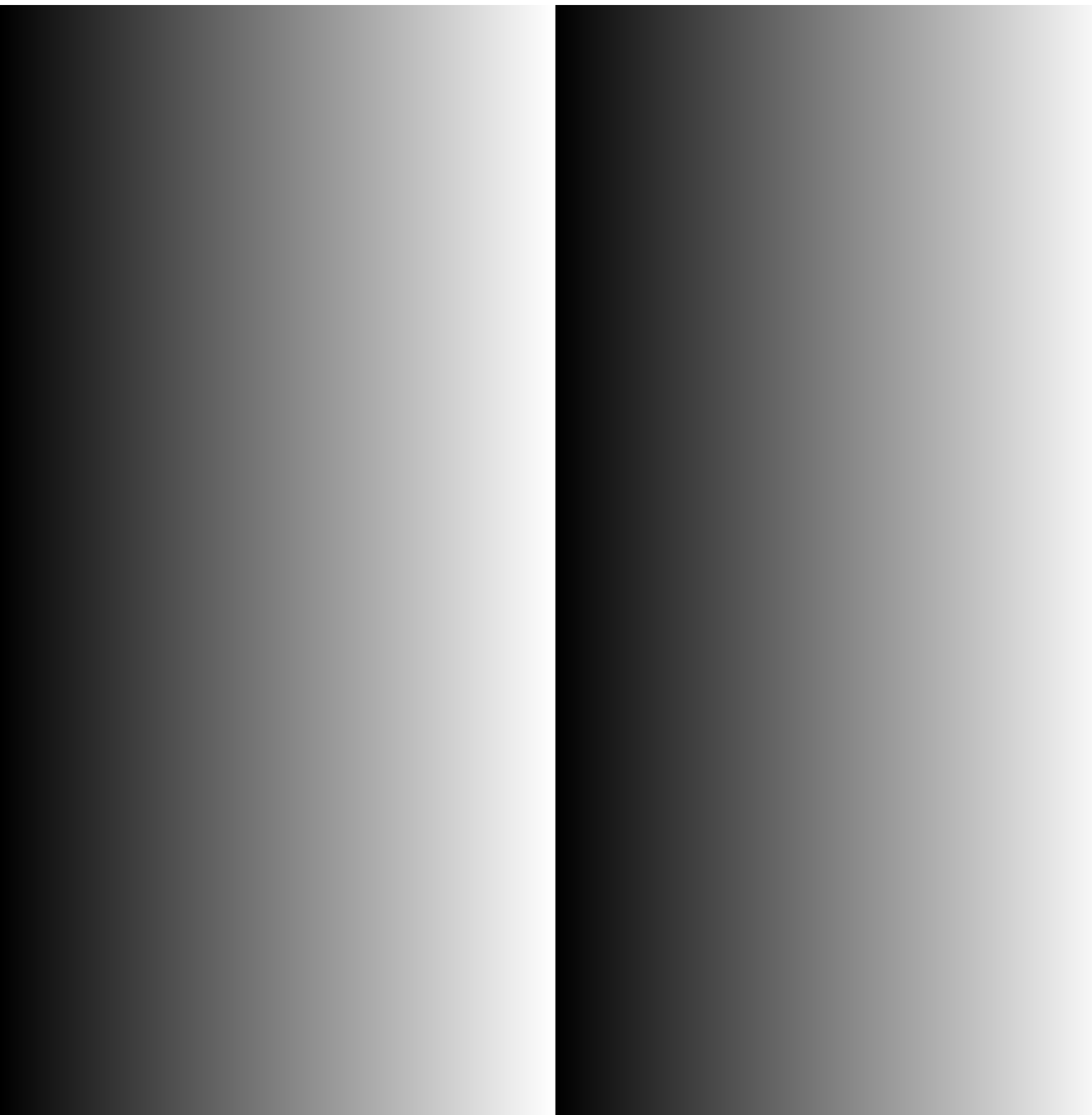}
b)
\end{minipage}
\begin{minipage}[t]{0.75\imagewidth}
\raggedright
\includegraphics[width=0.75\imagewidth]{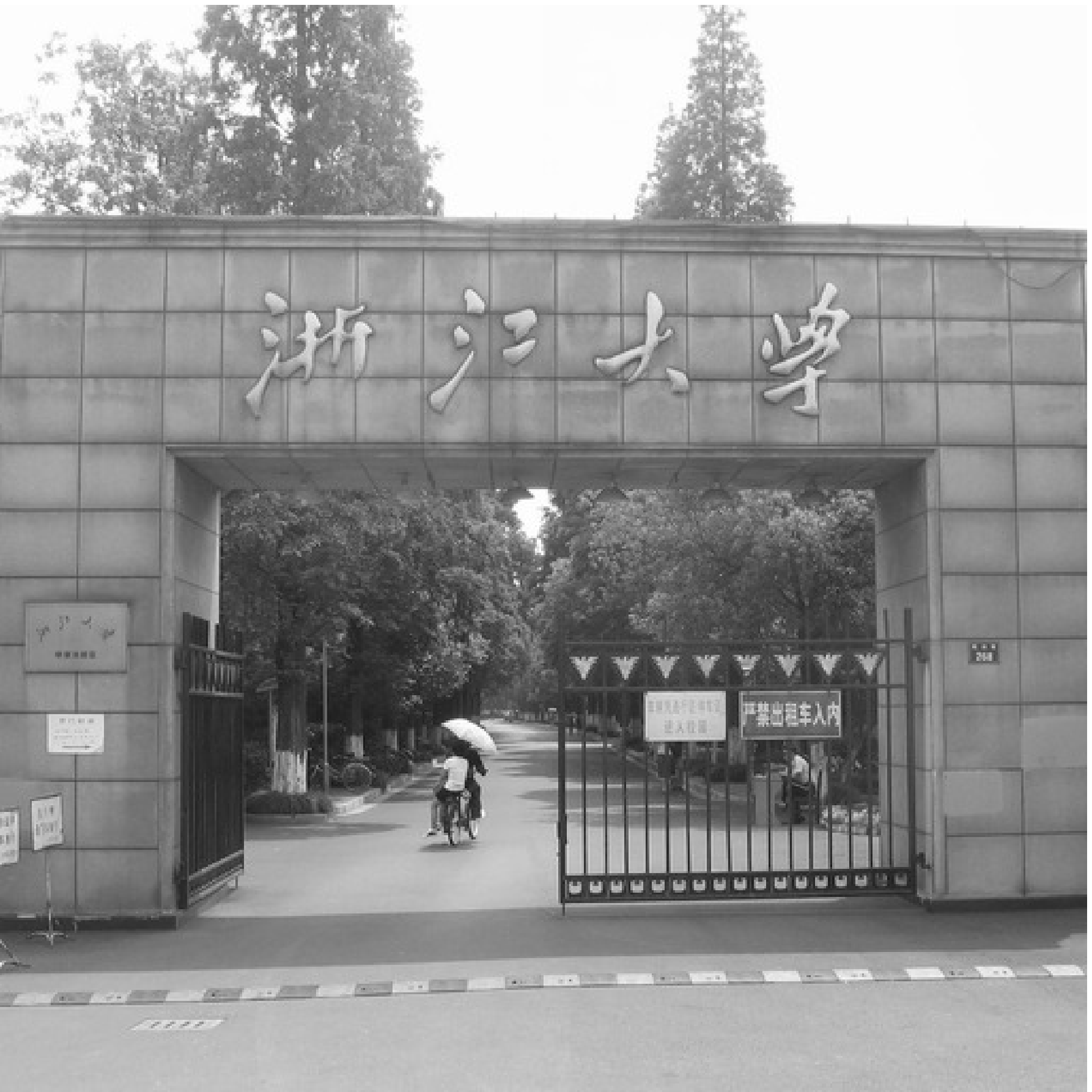}
c)
\end{minipage}
\centering
\begin{minipage}[t]{0.75\imagewidth}
\raggedright
\includegraphics[width=0.75\imagewidth]{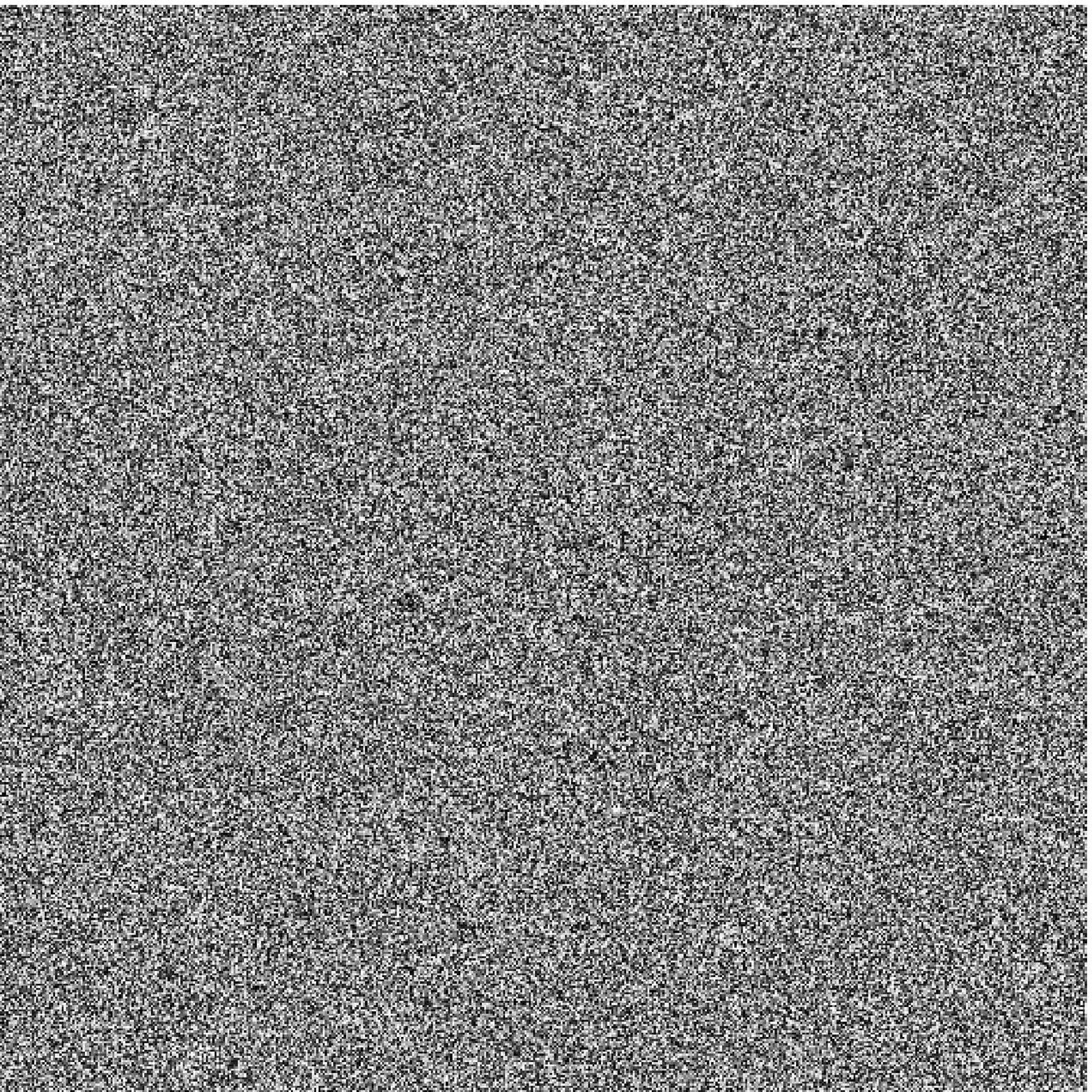}
d)
\end{minipage}
\begin{minipage}[t]{0.75\imagewidth}
\raggedright
\includegraphics[width=0.75\imagewidth]{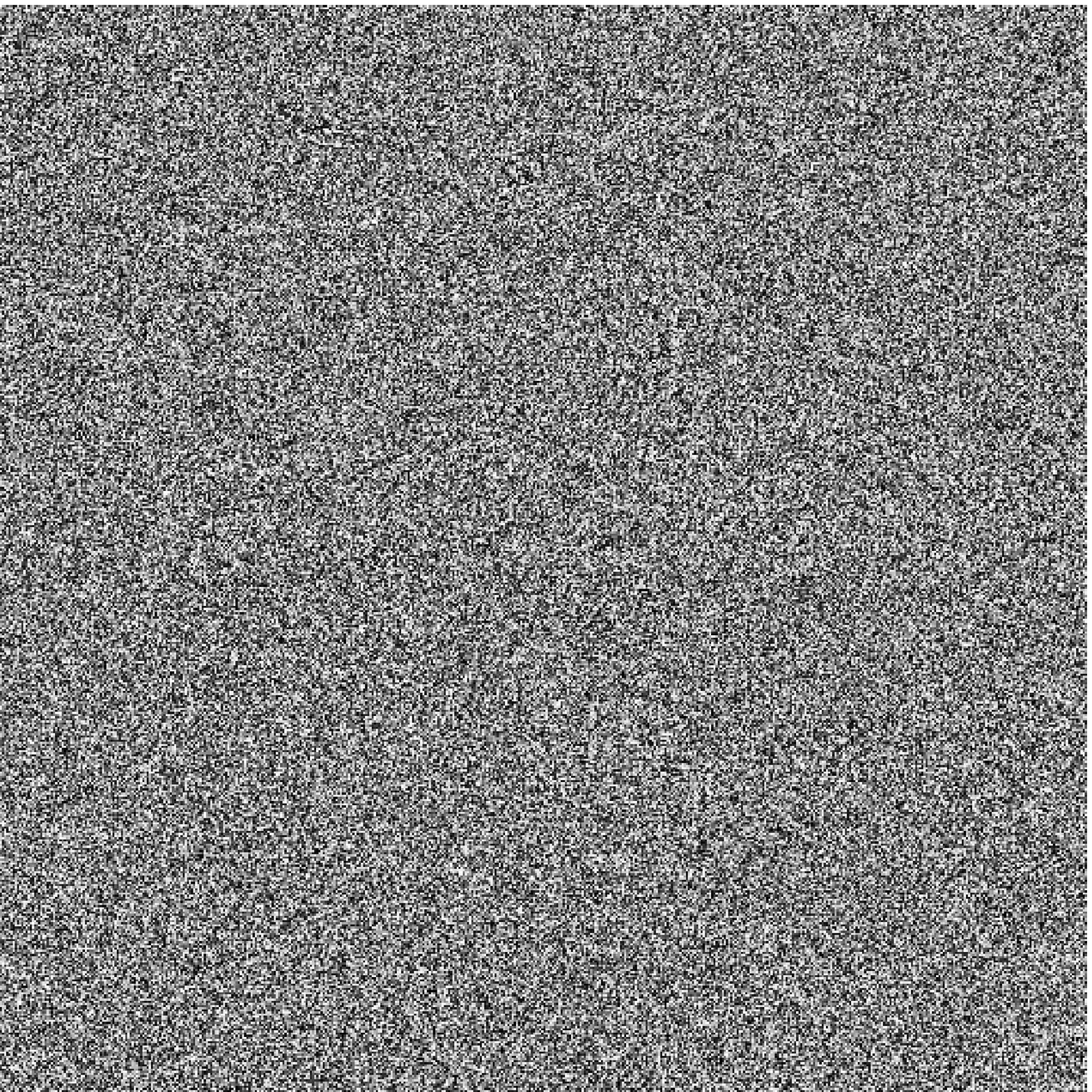}
e)
\end{minipage}
\begin{minipage}[t]{0.75\imagewidth}
\raggedright
\includegraphics[width=0.75\imagewidth]{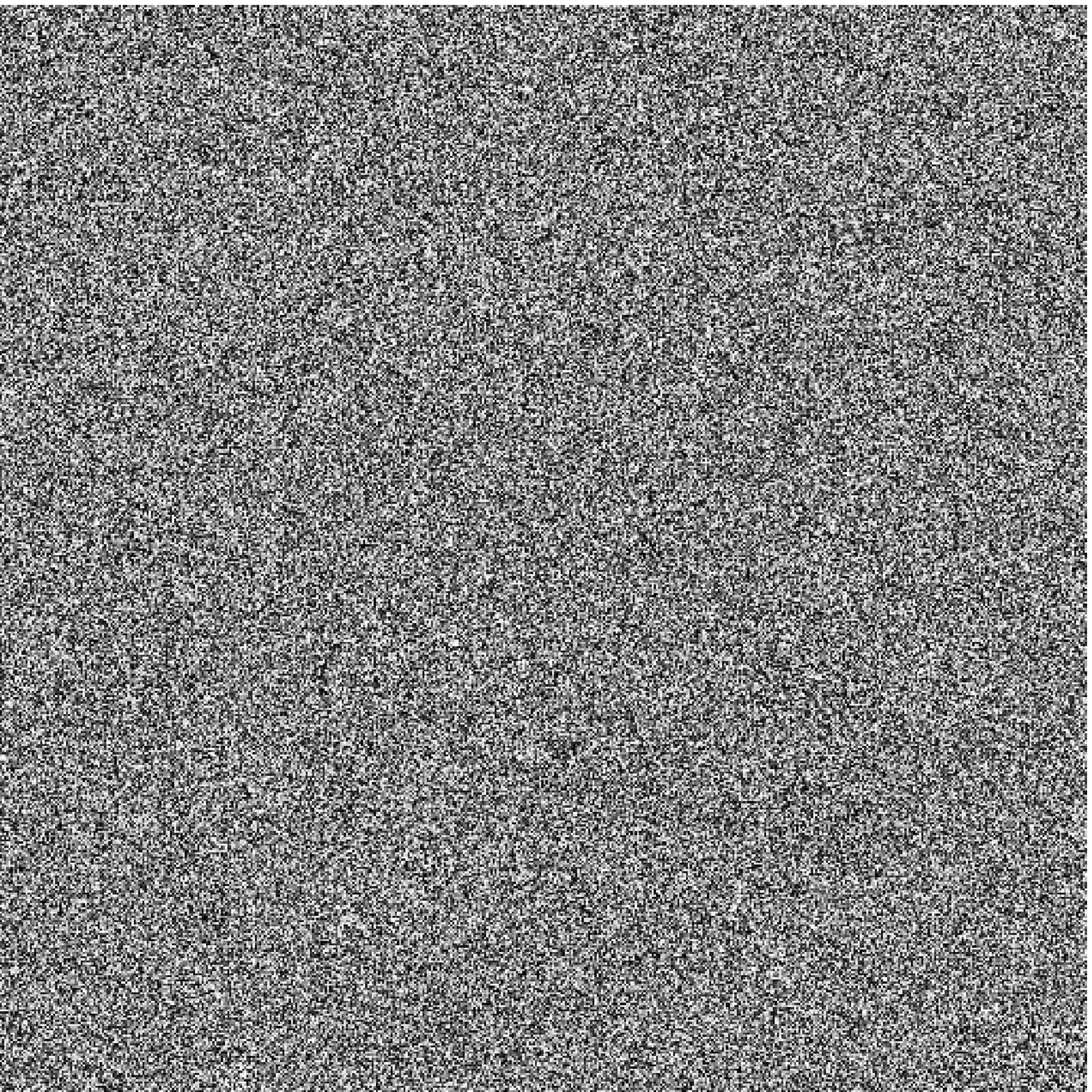}
f)
\end{minipage}
\caption{Three chosen plain-images for breaking position permutation
and the corresponding cipher-images.}
\label{figure:BreakingPermutation}
\end{figure}

\begin{figure}[!htb]
\centering
\begin{minipage}{0.8\figwidth}
\raggedright
\includegraphics[width=\textwidth]{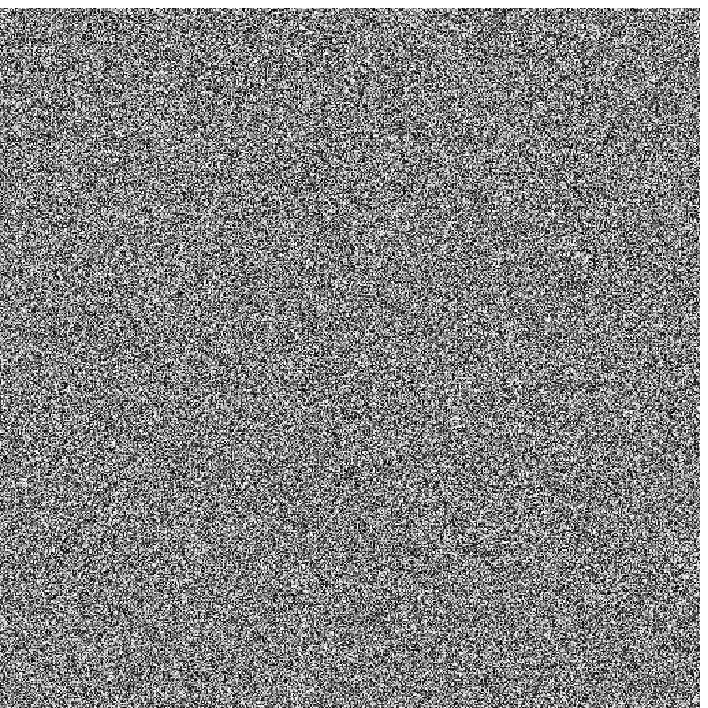}
a)
\end{minipage}
\begin{minipage}{0.8\figwidth}
\raggedright
\includegraphics[width=\textwidth]{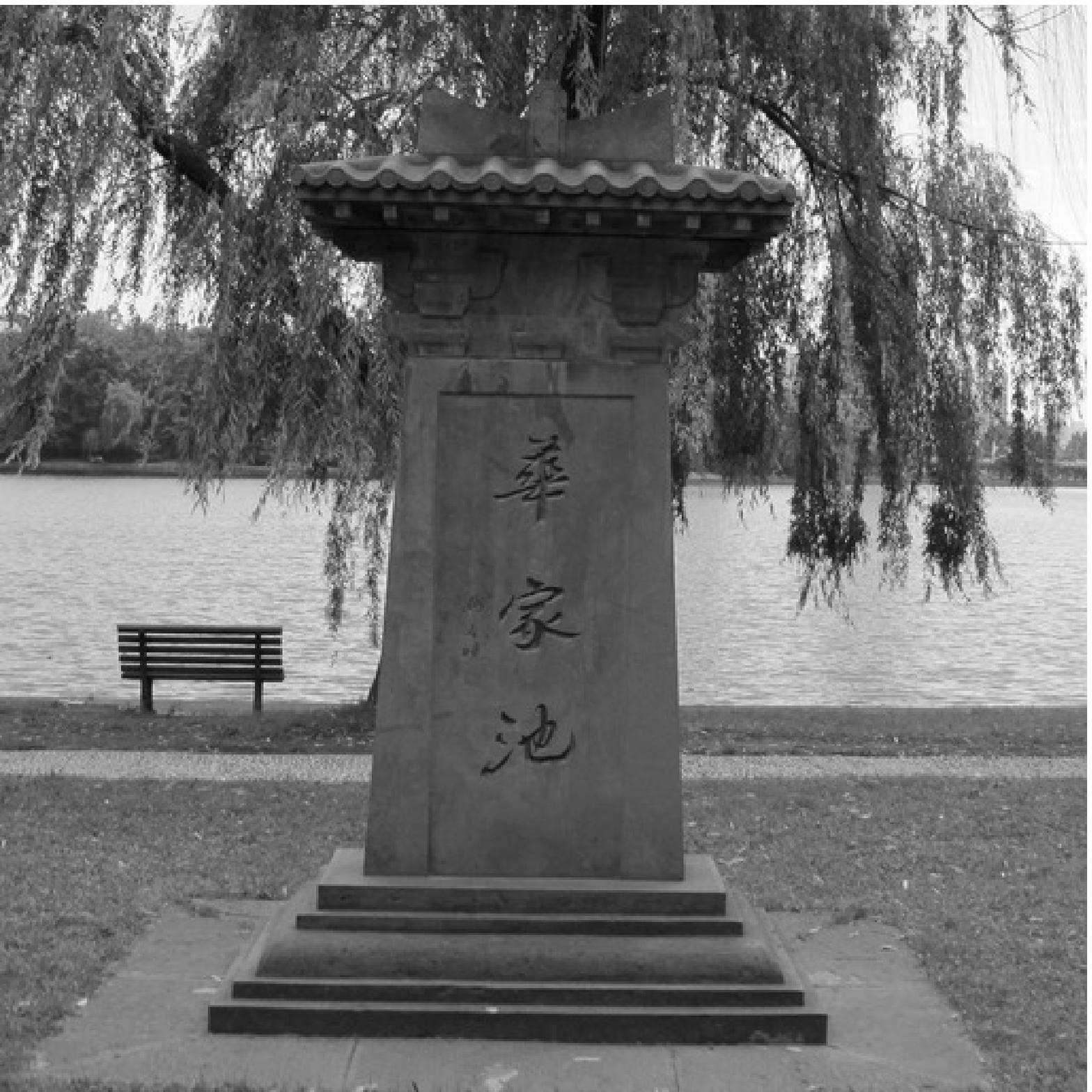}
b)
\end{minipage}
\caption{The recover of another plain-image encrypted with the same
secret key.}\label{figure:chosenplaintextattack}
\end{figure}

To validate performance of the proposed attack, some experiments on
some plain-images of size $512\times 512$ have been performed.
Besides $S=33$, the same secret key used in \cite[Sec.
3]{Behnia:IJBC08} was adopted: $(x_0, \alpha_1,
  \alpha_2)=(25.687, 2.10155, 3.569221)$, $(x_0', \alpha_1', \alpha_2')=(574.461, 1.8874, 4.23562)$,
  $(x_0^*, \alpha_1^*,  \alpha_2^*)=(814.217217, 2.8912, 3.89954 )$, $(y_0, \alpha_3,
  \alpha_4)=(79.82, 61.522, 257.26223)$. The step \textit{Confusion I} can be broken with the six chosen
plain-images of fixed values, 9, 127, 1, 52, 33 and 65, as shown in
Fig.~\ref{figure:BreakingConfusion} and the corresponding
cipher-images shown in Fig.~\ref{figure:BreakingConfusion2}. Then,
the plain-image shown in Fig.~\ref{figure:BreakingConfusion}a) and
the corresponding cipher-images can break the step \textit{Confusion
II}. Finally, the step \textit{Permutation} can be broken with
$\lceil \log_{256}(512\cdot 512)\rceil =3$ special plain-images
shown in Fig.~\ref{figure:BreakingPermutation}. The obtained
equivalent secret key was used to decrypt another cipher-image, as
shown in Fig.~\ref{figure:chosenplaintextattack}a), and the result
is shown in Fig.~\ref{figure:chosenplaintextattack}b).

\subsection{Some Other Security Defects}

\begin{itemize}
\item Problems about Secret Key;

As specified in \cite[Rule 5]{LiShujun:Rules:IJBC2006}, the key
space of a secure encryption scheme should be precisely specified
and avoid non-chaotic regions. However, even with the measure used
in \cite{Behnia:IJBC08}, a great number of secret key should be
excluded from the key space of the encryption scheme under study
(see Fig.~\ref{figure:positiveexponentf(x)}).

\begin{figure}[!htb]
\centering
\begin{minipage}{\figwidth}
\centering
\includegraphics[width=\textwidth]{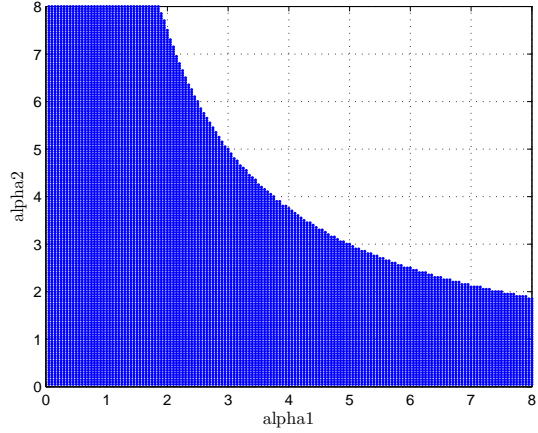}
\end{minipage}
\caption{The parameters of $f(x)$ corresponding to positive Lyapunov
exponent.} \label{figure:positiveexponentf(x)}
\end{figure}

\item Insufficient Randomness of Pseudo-Random Number Sequences
$\{\phi_1(k)\}$, $\{\phi_2(k)\}$, $\{\phi_3(k)\}$, and
$\{\phi_4(k)\}$

To study the dynamic property of the two equations $f(x)$ and
$g(x)$, we drew the graph of the two equations under a greater of
number of random parameters. Due to the similarity, only the graphs
of $f(x)$ and $g(x)$ with $(\alpha_1, \alpha_2)=(2.10155, 3.56922)$,
$(\alpha_3, \alpha_4)=(61.522,257.26223)$ are shown in
Fig.~\ref{figure:graphf(x)g(x)}. Comparing the graphs of the two
functions and $y=x$, one can assure that the states generated by
iterating the two functions will approach zero soon after some
iterations.

\begin{figure}[!htb]
\centering
\begin{minipage}{\figwidth}
\centering
\includegraphics[width=\figwidth,height=\figwidth]{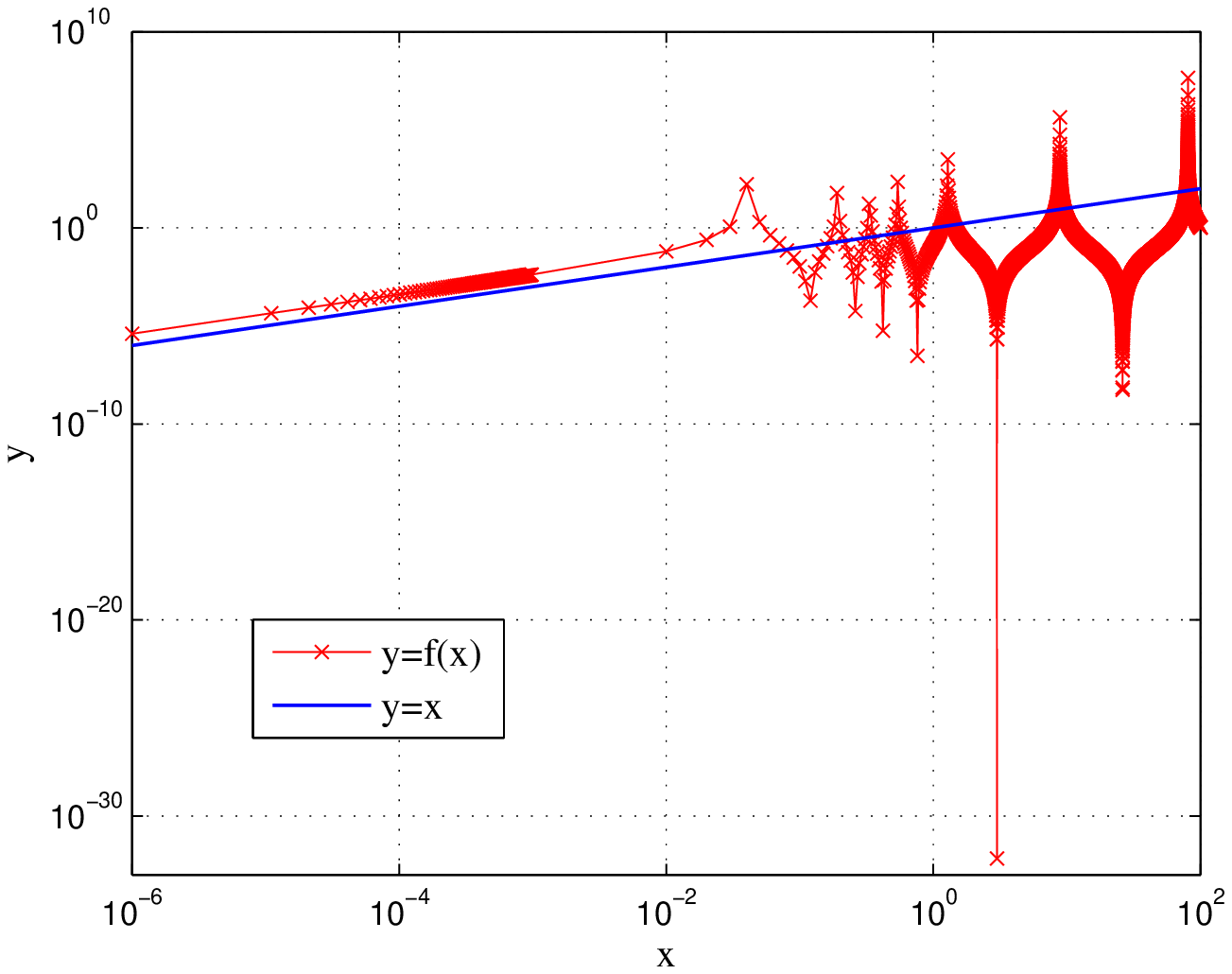}
a) $f(x)$
\end{minipage}
\begin{minipage}{\figwidth}
\centering
\includegraphics[width=\figwidth,height=\figwidth]{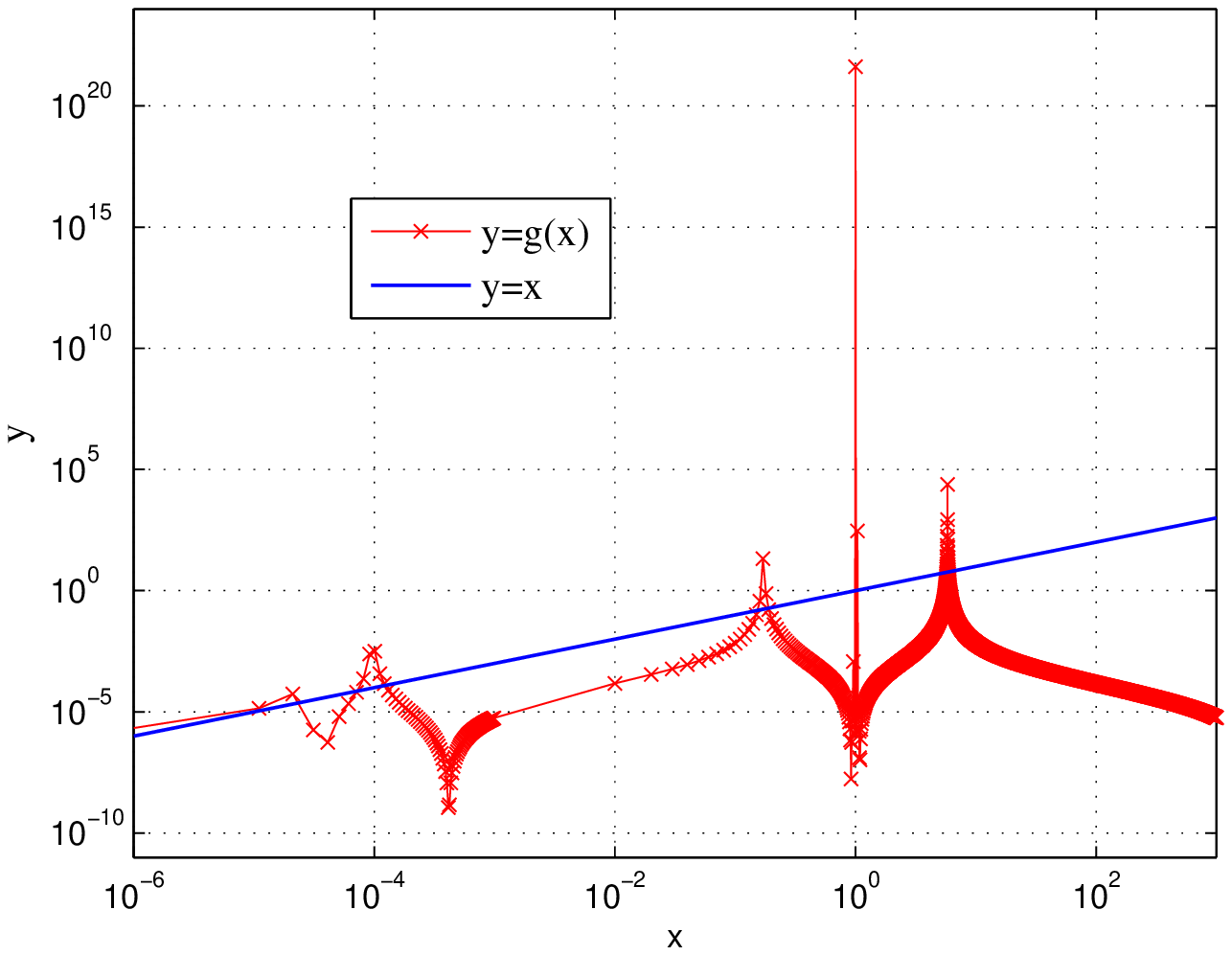}
b) $g(x)$
\end{minipage}
\caption{The graph of $f(x)$ and
$g(x)$.}\label{figure:graphf(x)g(x)}
\end{figure}

To further test the randomness of the sequences generated by the two
equations, we adopted the test suite proposed in
\cite{Rukhin:TestPRNG:NIST}. Since the three sequences
$\{\phi_1(k)\}$, $\{\phi_2(k)\}$, and $\{\phi_4(k)\}$ are determined
by the same equation, only the randomness of $\{\phi_3(k)\}$ and
$\{\phi_4(k)\}$ was tested. For every sequence, 100 samples of
length $512\cdot 512/8=32768$ (the number of bytes used for
encryption of a gray-scale plain-image of size $512\times 512$) were
generated by random secret keys. For each test, the default
significance level 0.01 was adopted. The results are shown in
Table~\ref{table:test}, from which one can see that the two
equations both cannot be used as a good random number generator.

\begin{table}[!htbp]
\centering\caption{The performed tests with respect to a
significance level 0.01 and the number of sequences passing each
test in 100 randomly generated sequences.}\label{table:test}
\begin{tabular}{c|c|c}
\hline
\multirow{2}{1in}{Name of Test}     & \multicolumn{2}{c}{Number of Passed Sequences}\\
\cline{2-3}                                            & $f(x)$ & $g(x)$\\
\hline\hline Frequency                                 & 6      & 2 \\
\hline Block Frequency ($m=100$)                       & 10     & 6  \\
\hline Cumulative Sums-Forward                         & 6      & 3 \\
\hline Runs                                            & 8      & 3  \\
\hline Rank                                            & 68     & 99  \\
\hline Non-overlapping Template ($m=9$, $B=110001000$) & 76     & 64 \\
\hline Serial ($m=16$)                                 & 6      & 9  \\
\hline Approximate Entropy ($m=10$)                    & 8      & 6  \\
\hline FFT                                             & 65     & 49  \\
\hline
\end{tabular}
\end{table}

\item Insensitivity with Respect to Changes of Plaintext

In \cite[Sec.5.4]{Behnia:IJBC08}, the importance of sensitivity with
respect to changes of plaintext is recognized. However, the
encryption scheme under study is actually very far away from the
desired property. In cryptography, the most ideal situation about
sensitivity is that the change of any single bit of plaintext will
make every bit of the corresponding ciphertext change with a
probability of one half. Obviously, the encryption scheme under
study can not reach the desired state due to the following points.

\begin{itemize}
\item No nonlinear S-box is involved in the whole scheme;

\item Any bit of plaintest only may influence the bits at the above levels in
the ciphertext;

\item Any pixel of plaintext does not influence other pixels in
the corresponding ciphertext uniformly.
\end{itemize}

To demonstrate this defect efficiently, we performed an experiment
by changing a bit of the plain-image of size $512\times 512$ shown
in Fig.~\ref{figure:BreakingPermutation}c). It is found that only
the bits of one level are changed. The locations of the changed bits
are shown in Fig.~\ref{figure:sensitivity}, where the white dots
denote changed locations and black ones denote unchanged ones.

\begin{figure}[!htb]
\centering
\begin{minipage}[t]{0.6\imagewidth}
\raggedright
\includegraphics[width=0.6\imagewidth]{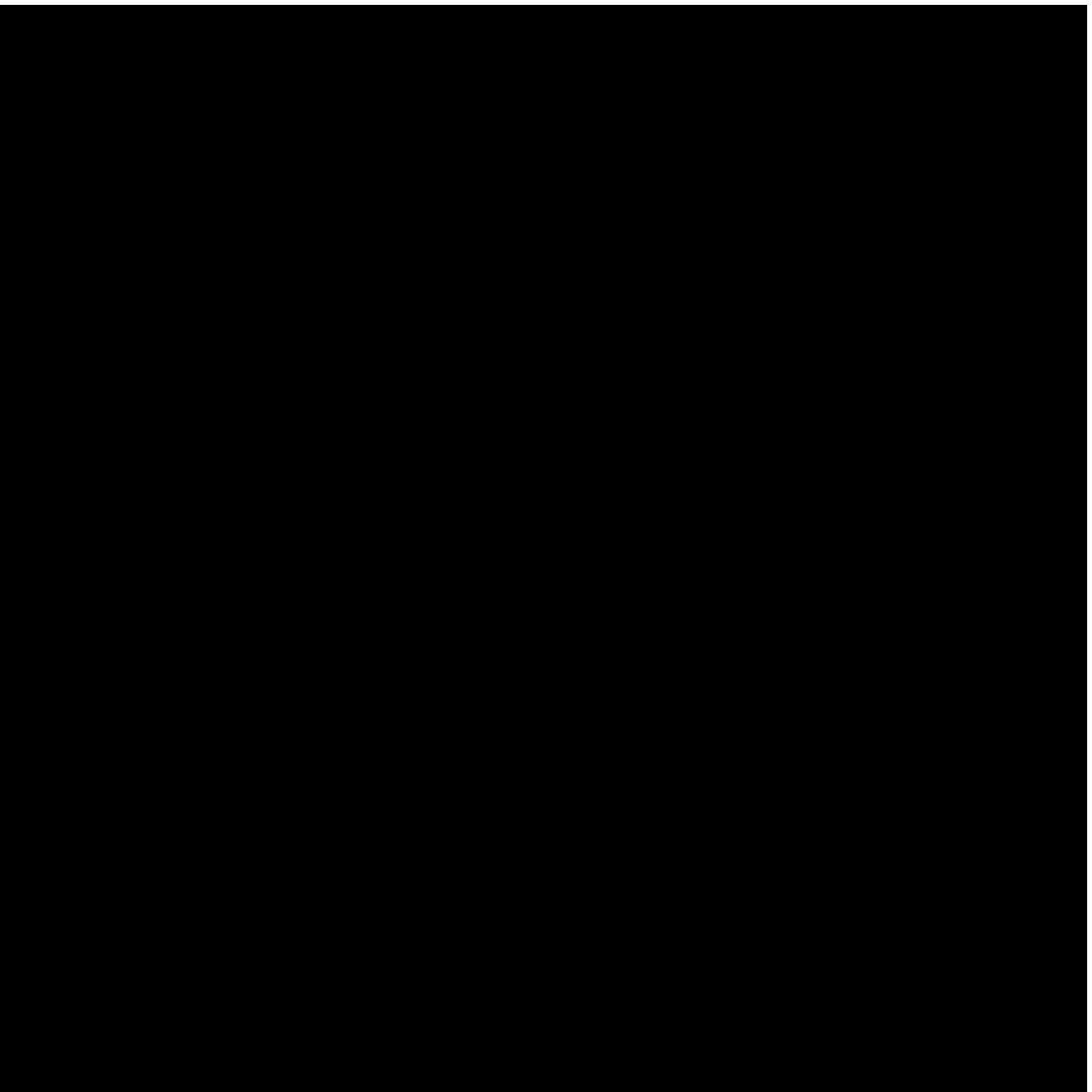}
a) $0\sim 4$th
\end{minipage}
\begin{minipage}[t]{0.6\imagewidth}
\raggedright
\includegraphics[width=0.6\imagewidth]{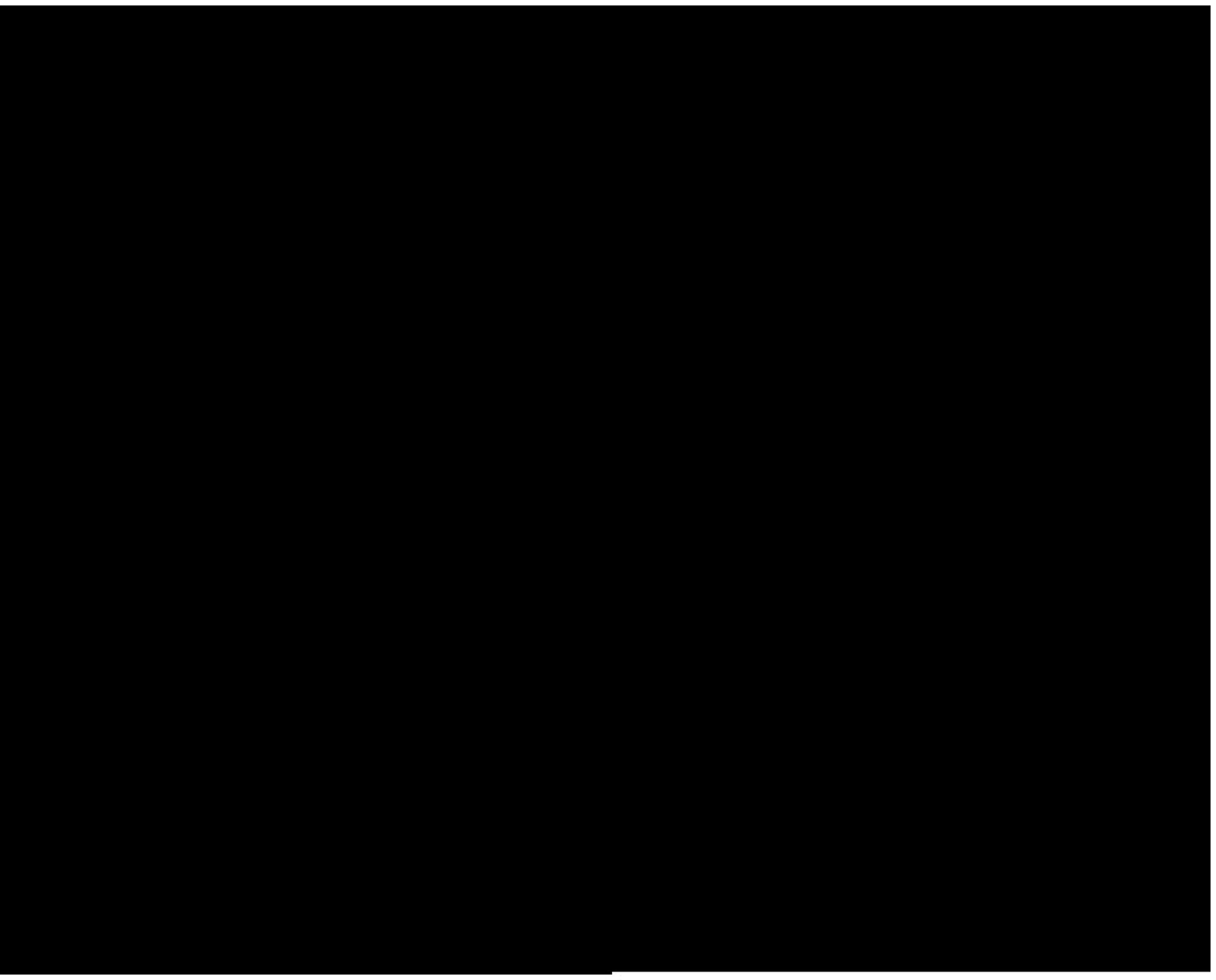}
b) 5th
\end{minipage}
\begin{minipage}[t]{0.6\imagewidth}
\raggedright
\includegraphics[width=0.6\imagewidth]{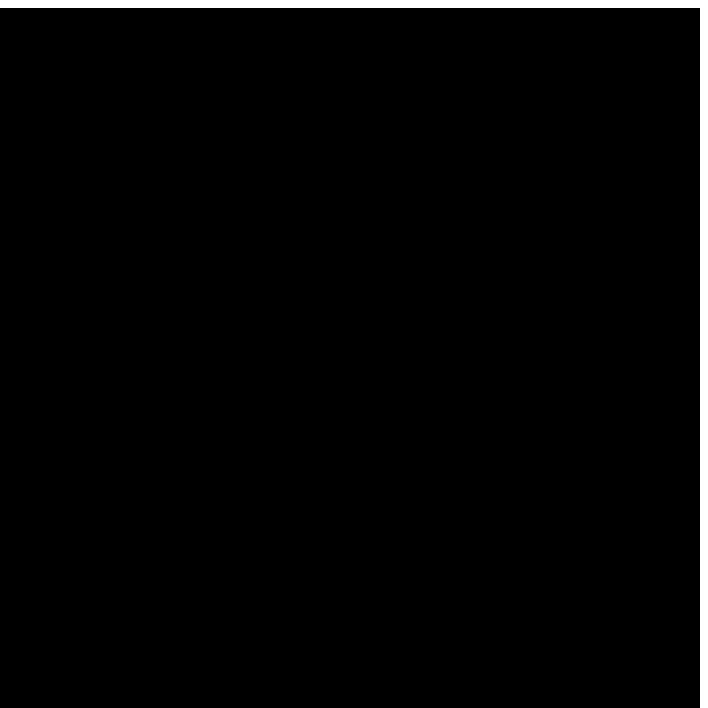}
c) 6th
\end{minipage}
\begin{minipage}[t]{0.6\imagewidth}
\raggedright
\includegraphics[width=0.6\imagewidth]{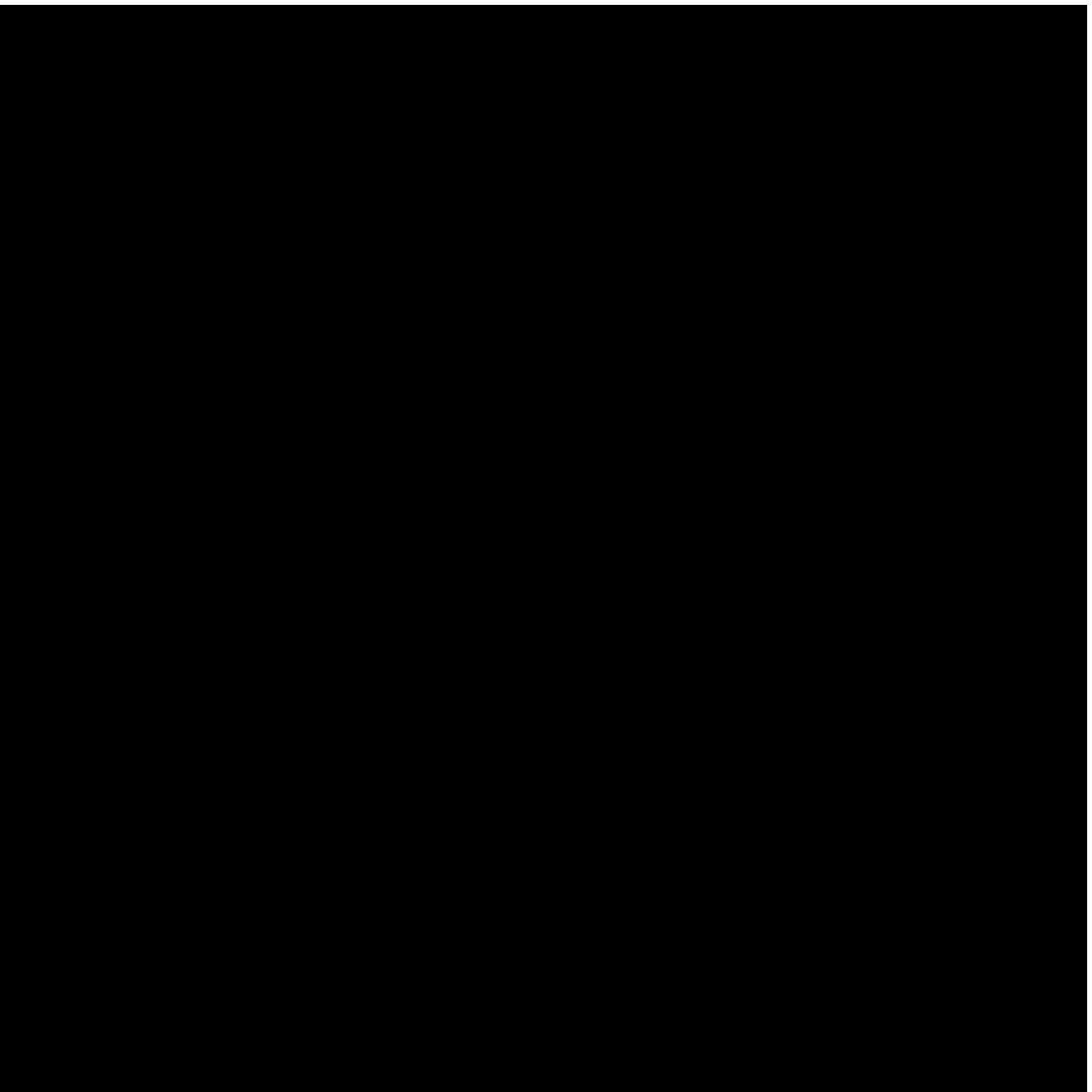}
c) 7th
\end{minipage}
\caption{The locations of changed bits of the cipher-image, when the
5-th bit of the pixel at location (256, 256) in the plain-image was
changed.}\label{figure:sensitivity}
\end{figure}
\end{itemize}

\section{Conclusion}
\label{sec:conclude}

In this paper, the security of a chaotic cryptographic scheme based
on composition maps has been studied in detail. It is found that the
scheme can be broken with $6+\lceil\log_L(MN)\rceil$
chosen-plaintext. In addition, the scheme is not sensitive to the
changes of plaintext also. Furthermore, the randomness of the
pseudo-random number sequences generated by the composition maps is
very weak. Due to the insecurity of the scheme under study, it is
should not be used in real serious application.

\section*{Acknowledgement}

Chengqing Li was supported by The Hong Kong Polytechnic University's
Postdoctoral Fellowships Scheme under grant no. G-YX2L. The work of
Kowk-Tung Lo was supported by the Research Grant Council of the Hong
Kong SAR Government under Project 523206 (PolyU 5232/06E).

\bibliographystyle{IEEEtran}
\bibliography{IJBC}

\end{document}